\documentclass[preprint,eqsecnum,showpacs,superscriptaddress,nofootinbib,aps,amsmath,amssymb,tightenlines]{revtex4}

\makeatother
\newcommand{\hateq}{\widehat{=}}

 \makeatletter
 \newcommand{\pback}[1]{{
   \let\@rrow=\leftarrowfill
\mathchoice{\AIN@stemPullBack{#1}{\@rrow}}{\AIN@stemPullBack{#1}{\@rrow}}
     {\AIN@indxPullBack{#1}{\@rrow}}{\AIN@indxPullBack{#1}{\@rrow}}}
   \vphantom{#1}}

 \newcommand{\AIN@stemPullBack}[2]{
   \vtop{\mathsurround=0pt
   \ialign{##\crcr$\textstyle{#1}\strut$\crcr
     \noalign{\kern-0.4ex\nointerlineskip}{\tiny#2}\crcr}}}

 \newcommand{\AIN@indxPullBack}[2]{
   \vtop{\mathsurround=0pt
   \ialign{##\crcr\hfil$\scriptstyle{#1}$\hfil\crcr
     \noalign{\kern+0.4ex\nointerlineskip}{\tiny#2}\crcr}}}

\makeatother

\def\ba{\begin{eqnarray}}
\def\ea{\end{eqnarray}}
\def\be{\begin{equation}}
\def\ee{\end{equation}}
\def\={\hateq}

\def\f{\frac}
\def\lp{\ell_{\rm Pl}}
\def\IH{\Delta}
\def\ub{\underbar}
\def\H{{\cal H}}
\def\SU(2){\rm SU(2)}
\def\U(1){\rm U(1)}
\def\g{\gamma}

\def\d{{\rm d}}
\preprint{\vbox{\baselineskip=12pt \rightline{gr-qc/0305082}
\rightline{CGPG-03-05/4}
 \rightline{ICN-UNAM-03-06}}}

\begin{document}

\title{Non-minimal couplings, quantum geometry\\ and black hole entropy}
\author{Abhay\ Ashtekar}\email{ashtekar@gravity.psu.edu}
\affiliation{Center for Gravitational Physics and Geometry Physics
Department, Penn State, University Park, PA 16802, USA}
\affiliation{Erwin Schr\"odinger Institute, Boltzmanngasse 9, 1090
Vienna, AUSTRIA}
\author{Alejandro\ Corichi}\email{corichi@nuclecu.unam.mx}
\affiliation{Instituto de Ciencias Nucleares, Universidad Nacional
Aut\'onoma de M\'exico, A. Postal 70-543, M\'exico D.F. 04510,
MEXICO}

\begin{abstract}
The black hole entropy calculation for type I isolated horizons,
based on loop quantum gravity, is extended to include
non-minimally coupled scalar fields. Although the non-minimal
coupling significantly modifies quantum geometry, the highly
non-trivial consistency checks for the emergence of a coherent
description of the quantum horizon continue to be met. The
resulting expression of black hole entropy now depends also on the
scalar field precisely in the fashion predicted by the first law
in the classical theory (with the same value of the
Barbero-Immirzi parameter as in the case of minimal coupling).
\end{abstract}


\maketitle
\section{Introduction}
\label{s1}

In classical general relativity, weakly isolated horizons provide
a unified framework to analyze properties of black hole and
cosmological horizons in equilibrium \cite{ihprl,afk,abl2,abl1},
where the geometry and matter fields on the horizon itself are
assumed to be time independent but the physics in the exterior
region can be dynamical. Although the horizons typically lie in a
highly curved region of space-time, their symmetry groups fall in
to three universality classes \cite{abl2}. Cases of direct
physical interest are the type I horizons where the
\emph{intrinsic} geometry and matter fields \textit{on} the
horizon are spherical and type II horizons where they are
axi-symmetric. Note that these symmetries need not extend in the
exterior region; a class of Robinson-Trautman solutions provide an
explicit example where the isolated horizon is of type I but where
the 4-geometry does not admit a Killing field in any neighborhood
of the horizon \cite{pc}.

The sector of general relativity consisting of space-times with a
weakly isolated horizon inner boundary admits an action principle
and a Hamiltonian formulation. It is therefore possible to carry
out canonical quantization of this sector. For type I horizons
this procedure was implemented in detail in \cite{abck,ack,abk}.
The implementation required an extension of the bulk quantum
geometry [9-22]
to accommodate the presence of a boundary, and the construction of
a $U(1)$ Chern-Simons theory to describe the geometry of the
quantum horizon. The requirement that the inner boundary is an
isolated horizon is incorporated in the quantum theory by
promoting the horizon boundary condition to an operator equation.
This allows the horizon to fluctuate but requires that the
fluctuations be correlated in a way dictated by the classical
boundary condition. The form of this \emph{quantum horizon
condition} is such that a coherent theory can emerge if and only
if eigenvalues of a certain operator in the quantum theory of the
bulk geometry are \emph{exactly} equal to those of another
operator in the surface Chern-Simons theory. This is a stringent
requirement because the two theories are quite independent and
eigenvalues of each operator can be computed in the respective
theory without any knowledge of the other! Yet, the boundary
conditions introduced in the isolated horizon framework relate the
parameters appearing in the two theories in just the right way for
the equality to hold (see section \ref{s2}).

Next, one can construct a micro-canonical ensemble by fixing the
horizon area and charges and calculate the number of microstates
in the ensemble. However, the bulk quantum geometry has a
1-parameter ambiguity, labelled by what is known as the
`Barbero-Immirzi parameter' $\gamma >0$. This has close similarity
with the $\theta$-ambiguity in QCD \cite{immirzi}. There are no
physical operators mixing states in distinct $\gamma$-sectors;
there is super-selection. As with the parameter $\theta$ in QCD,
the value of $\gamma$ in Nature is to be determined by
experiments. States and operators in various $\gamma$ sectors are
very similar in their structure but the eigenvalues of geometrical
operators scale with $\gamma$. Hence the $\gamma$ ambiguity
trickles down to the expression of the number of horizon states.
Irrespective of the value of $\gamma$, the entropy turns out to be
proportional to the horizon area $a_o$. However, the coefficient
depends on the value of $\gamma$. Since there is a single
undetermined parameter, its value can be fixed by a single
`experiment' ---for example, by measuring the smallest eigenvalue
of the area operator. Unfortunately, technology necessary for a
direct measurement of this type is not available. But we can use
the Bekenstein-Hawking semi-classical entropy formula
\be \label{S1} S = \f{1}{4\,\lp^2}\, a_o\, , \ee
where, $\lp$ is the Planck length, to `carry out an indirect
measurement'. Suppose, for example, that we demand that the
leading term in the expression of entropy of a large Schwarzschild
black hole be given by (\ref{S1}). This fixes the value of
$\gamma$,
\be \label{bi} \gamma = \f{\ln 2}{\pi\sqrt{3}}\,  \ee
and hence the theory. One can now test this theory. In particular,
is the entropy of charged black holes or of cosmological horizons
correctly recovered in this theory? In \cite{abk}, the answer was
shown to be in the affirmative for type I  ---i.e., non-rotating,
undistorted---  horizons. That work has since been extended to
type II horizons, i.e., to incorporate rotation and distortion
\cite{aaentropy}.

These calculations incorporated the possible presence of Maxwell
and scalar fields at the horizon (possibly with dilatonic
couplings), allowing for non-zero electric, magnetic and dilatonic
charges. However all these fields are \emph{minimally coupled to
gravity}. Now, using Killing horizons, Jacobson, Kang and Myers
\cite{jkm} and Iyer and Wald \cite{iw} showed that in presence of
\emph{non-minimal} couplings to gravity, the first law of black
hole mechanics in classical general relativity is non-trivially
modified, suggesting that the entropy should now depend not only
on the area but also on the values of matter fields at the
horizon. For non-minimally coupled scalar fields, this analysis
was recently extended to the isolated horizon framework
\cite{acs2}. Specifically, in the theory
governed by the action:%
\footnote{Here we have ignored the surface term. In loop quantum
gravity, one uses a first order framework based on tetrads and
connections. The first order action for this theory, including the
surface term, is given in \cite{acs2}.}
\be
 {\cal S}[ g_{ab}, \phi] =\int \d^4\!x \sqrt{- g}\left[
\frac{1}{16\pi G}\, f(\phi) R - \frac{1}{2}  g^{ab} \,\partial_a
\phi\, \partial_b \phi - U(\phi)\right]\, , \label{action} \ee
where $R$ is the scalar curvature of the metric $g_{ab}$ and $U$
is a potential for the scalar field, the entropy is given by
\cite{acs2}
\be \label{S2} S = \f{1}{4\,\lp^2}\, \oint_S f(\phi)\, \d^2 V \ee
where $S$ is any 2-sphere cross-section of the horizon. A natural
question now is: Can the quantum geometry calculation incorporate
this situation? At first sight, this seems to be difficult because
matter fields at the horizon play no essential role in that
calculation; the calculation is dictated almost entirely by the
geometry of the quantum horizon.

Now, in the case when $f$ is nowhere vanishing, one could first
re-express the classical theory using the `Einstein frame' by an
appropriate conformal rescaling of the metric that removes the
non-minimal coupling and \emph{then} carry out quantization as in
\cite{ack,aaentropy}. However, that procedure would simply
`by-pass' the issue, rather than meeting it `head-on', leaving the
ramifications of non-minimal coupling unexplored. In the classical
theory, one can carry out the entire analysis in either the
non-minimally coupled Jordan frame or minimally coupled Einstein
frame and \emph{demonstrate} that results agree \cite{acs2}. Can
one do the same in the quantum theory? Already at the classical
level, a priori, it is not obvious that the agreement must hold
(see section V of \cite{acs2}). At the quantum level, it is even
less clear that the stringent requirements for the emergence of a
coherent description of the quantum horizon can be satisfied in
the Jordan frame. Finally, the derivation of the first law in
\cite{acs2} was carried out using a first order action which is
equivalent to (\ref{action}) if $f$ is nowhere vanishing. However,
in the first order formalism $f$ can be allowed to vanish on open
sets which remain bounded away from the horizon and infinity; the
first law still holds, with entropy given by (\ref{S2}). In the
first order framework, in general it is not even possible to pass
to the Einstein frame and a direct analysis in the Jordan frame is
necessary.

In this paper, for simplicity, we will restrict ourselves to type
I isolated horizons and confront the non-minimal coupling directly
in the Jordan frame using the first order framework of
\cite{acs2}. We will find that the presence of non-minimal
coupling introduces a major modification in the quantum theory of
the bulk geometry and also changes the `level' (i.e. the coupling
constant) of the surface Chern Simons theory. But the two
modifications conspire to leave the delicate matching between the
bulk and horizon quantum structures in tact, whence a coherent
theory of the geometry of the quantum horizon continues to exist
also in the Jordan frame. One can then calculate entropy. One now
finds that for large black holes the entropy is given by
(\ref{S2}) (rather than (\ref{S1})) for the same value of the
Barbero-Immirzi parameter $\gamma$ as in the case of the minimal
coupling.

\section{Minimally coupled matter: Summary}
\label{s2}

To stream-line the calculation and to bring out the modifications
brought about by non-minimal coupling, we will first recall the
highlights of the analysis in the case when all fields are
minimally coupled to gravity. For brevity, we will overlook
subtleties, some of which are conceptually important. These are
discussed in detail in \cite{abk}.

In loop quantum gravity, one begins with a Hamiltonian formulation
of general relativity. The configuration variable $A_a^i$ is an
$\SU(2)$ connection on a 3-manifold $M$ and the momentum variable
is represented by a 2-form field $\Sigma_{ab}^{i}$ which takes
values in the Lie algebra of $\SU(2)$. $A_a^i$ is a
spin-connection on $M$ and $E^a_i := \gamma\, \eta^{abc}
\Sigma_{bc \,i}$ has the physical interpretation of an orthonormal
triad of density weight $1$, where $\gamma >0 $ is the
Barbero-Immirzi parameter and $\eta^{abc}$ is the metric
independent, density weighted Levi-Civita 3-form on $M$.

Let us focus on the sector of the theory consisting of space-times
which admit a type I isolated horizon $\Delta$ with a fixed area
$a_o$ as the internal boundary. Then $M$ is asymptotically flat
and has an internal boundary $S$, topologically a  2-sphere, the
intersection of $M$ with $\Delta$. Introduce on $S$ an internal,
unit, radial vector field $r^i$ (i.e. any isomorphism from the
unit 2-sphere in the Lie algebra of $\SU(2)$ to $S$). Then it
turns out that \emph{the intrinsic geometry of $S$ is completely
determined by the pull-back $\ub{A}^ir_i =: W$ to $S$ of the
(internal-radial component of the) connection $A^i$ on $M$}
\cite{abl1}. Furthermore, $W$ is in fact a spin-connection
intrinsic to the 2-sphere $S$. Finally, the fact that $S$ is (the
intersection of $M$ with) a type I isolated horizon is captured in
a relation between the two canonically conjugate fields:
\be \label{bc1} F:\=\,\, \d W\,\, \=  - \f{2\pi\g}{a_o}\,\,
\underline\Sigma^i\, r_i .\ee
where $\underline{\Sigma}^i$ is the pull-back to $S$ of the
2-forms $\Sigma^i$ on $M$. (Throughout, $\=$ will stand for
equality restricted to $\Delta$.) Thus, because of the isolated
horizon boundary conditions, fields which would otherwise be
independent are now related. In particular, the pull-backs to $S$
of our canonically conjugate fields $A^i,\, \Sigma^i$ are
completely determined by the $U(1)$ connection $W$.

In absence of an internal boundary, the symplectic structure is
given just by a volume integral \cite{aa}. In presence of the
internal boundary under consideration, it now acquires a surface
term \cite{ack,abk}:
\be \label{sym1} {\bf \Omega}(\delta_1, \delta_2) = \f{1}{8\pi
G}\left[ \int_M \, {\rm  Tr}\, (\delta_1 A \wedge \delta_2 \Sigma
- \delta_2 A \wedge \delta_1 \Sigma) \, +\, \f{a_o}{\g \pi}
\oint_S \, \delta_1 W \wedge \delta_2 W\,   \right]\, , \ee
where $\delta \equiv (\delta A, \delta \Sigma)$ denotes tangent
vectors to the phase space ${\bf \Gamma}$. Since $W$ is
essentially the only `free data' on the horizon, it is not
surprising that the surface term of the symplectic structure is
expressible entirely in terms of $W$. However, it is interesting
that the new surface term is precisely the symplectic structure of
the $\U(1)$-Chern Simons theory. \emph{The symplectic structures
of the Maxwell, Yang-Mills, scalar and dilatonic fields do not
acquire surface terms and, because of minimal coupling, do not
feature in the gravitational symplectic structure either.}
Conceptually, this is an important point: this, in essence, is the
reason why the black hole entropy depends just on the horizon area
and not, in addition, on the matter charges \cite{abk}.

One can systematically `quantize' this sector of the phase space
\cite{abk}. We can focus only on the gravitational field since the
matter fields do not play a significant role. One begins with a
Kinematic Hilbert space $\H = \H_{V}\otimes \H_{S}$ where $\H_V$
is the Hilbert space of states in the bulk
\cite{al1,jb1,al2,mm,al3} and $\H_S$ is the Hilbert space of
surface states. Expression (\ref{sym1}) of the symplectic
structure implies that $\H_S$ should be the Hilbert space of
states of the Chern-Simons theory on the punctured $S$, where the
`level', or the coupling constant, is given by:
\be \label{level1} k = \f{a_o}{4\pi\gamma\lp^2} \ee
A pre-quantization consistency requirement is that $k$ be an
integer \cite{abk}.

Our next task is to encode in the quantum theory the fact that
$\Delta$ is a type I horizon with area $a_o$. This is done by
imposing the horizon boundary condition (\ref{bc1}) as an
\emph{operator equation}:
\be \label{qbc1} (1\otimes \hat{F})\, \Psi\, \=\, -
\left(\frac{2\pi\gamma}{a_o}\, (\hat{\underline{\Sigma}}\cdot
r)\otimes 1\right)\, \Psi \, , \ee
on admissible states $\Psi$ in $\H$. Now, a general solution to
(\ref{qbc1}) can be expanded out in a basis: $\Psi = \sum_n\,
\Psi_V^{(n)} \otimes \Psi_S^{(n)}$, where $\Psi_V^{(n)}$ is an
eigenvector of the `triad operator' $- ({2\pi\gamma}/{a_o})\,
(\hat{\underline{\Sigma}}\cdot r)(x)$ on $\H_V$ and $\Psi_S^{(n)}$
is an eigenvector of the `curvature operator' $\hat{F}(x)$ on
$\H_S$ \emph{with same eigenvalues}. Thus, the theory is
non-trivial only if a sufficiently large number of eigenvalues of
the two operators coincide. Since the two operators act on
entirely different Hilbert spaces and are introduced quite
independently of one another, this is a \emph{very} non-trivial
requirement.

Now, in the bulk Hilbert space $\H_V$, the eigenvalues of the
`triad operator' are given by \cite{al4}:
\be  \label{triadev1} -\, \left(\f{2\pi\gamma}{a_o}\right)\,\,
\left(8\pi \lp^2\,\, \sum_I\, m_I \delta^3(x, p_I)\,
\eta_{ab}\right)\, , \ee
where $m_I$ are half integers and $\eta_{ab}$ is the natural,
metric independent Levi-Civita density on $S$ and $p_I$ are points
on $S$ at which the polymer excitations of the bulk geometry in
the state $\Psi_V$ puncture $S$. A completely independent
calculation \cite{abk}, involving just the surface Hilbert space
$\H_S$, yields the following eigenvalues of $\hat{F}(x)$:
\be \label{Fev1} \f{2\pi}{k}\, \sum_I\, n_I\, \delta^3(x, p_I)\,
\equiv \,  2\pi\, \f{4\pi \gamma\lp^2}{a_o}\, \sum_I\, n_I
\,\delta^3(x, p_I)\, \ee
where $n_I$ are integers modulo $k$. Thus, with the identification
$-2m_I = n_I\, {\rm mod}\, k$, the two sets of eigenvalues match
exactly. Note that in the Chern-Simons theory the eigenvalues of
$F(x)$ are dictated by the `level' $k$ and the isolated horizon
boundary conditions tie it to the area parameter $a_o$ just in the
way required to obtain a coherent description of the geometry of
the quantum horizon.

In the classical theory, the parameter $a_{o}$ in the expression
of the surface term of the symplectic structure (\ref{sym1}) and
in the boundary condition (\ref{bc1}) is the horizon area. However
in the \emph{quantum theory}, $a_{o}$ has simply been a parameter
so far; we have not tied it to the \emph{physical area of the
horizon}. Therefore, in the entropy calculation, to capture the
intended physical situation, one constructs a suitable
`micro-canonical' ensemble. This leads to the last essential
technical step.

Let us begin by recalling that, in quantum geometry, the area
eigenvalues are given by \cite{rs3,al4}
$$ 8\pi \gamma\lp^2\, \sum_I \sqrt{j_I(j_I +1)}\, . $$
We can therefore construct a micro-canonical ensemble by
considering only that sub-space of the volume theory which, at the
horizon, satisfies:
\be \label{micro1} a_{o} -\epsilon \le 8\pi\gamma \lp^2\, \sum_I\,
\sqrt{j_I(j_I+1)} \le a_{o} + \epsilon \ee
where $I$ ranges over the number of punctures, $j_I$ is the spin
label associated with the puncture $p_I$ \cite{al4,abk}.%
\footnote{The appearance of the parameter $\epsilon$ is standard
in statistical mechanics. It has to be much smaller than the
macroscopic parameters of the system but larger than level
spacings in the spectrum of the operator under consideration. Its
precise value is irrelevant and does not affect the leading
contribution to entropy. We require $8\sqrt{3}\pi \gamma\lp^2 <
\epsilon \ll a_o$.}
In presence of matter fields carrying charges, we fix values of
horizon charges $Q_{o}$ and restrict the bulk matter states so
that
\be Q_{o} -\epsilon^\prime \le Q_{\rm hor} \le Q_{o} +
\epsilon^\prime \ee
for suitably chosen $\epsilon^\prime$s (one for each charge).
Finally, to obtain entropy, we have to calculate the number of
\emph{surface states} in this ensemble, i.e., in the sub-space of
$\H$ in which the quantum boundary conditions and Einstein's
equations are satisfied and in which the bulk states satisfy the
condition spelled out in (\ref{micro1}). The number is given by:
\be \label{N} {\cal N} = \sum_{p}\,\, \sum_{j_1,\ldots, j_p}\,\,
   \prod_{I=1}^p \,\, (2j_I +1) \ee
where the number $p$ of punctures and the spin-labels $j_1, ...
j_p$ are chosen to satisfy the area constraint above. Through
detailed analysis \cite{abk}, one can estimate the right side of
(\ref{N}) and calculate the entropy of large black holes:
\be \label{S3} S_{\IH} := \ln {\cal N} = \frac{\gamma_o}{\gamma}\,
\frac{a_{o}}{4\lp^2} + o \left(\frac{\lp^2}{a_{o}}\right) ,\quad
{\rm where} \quad \gamma_o = \frac{\ln 2}{\pi \sqrt{3}  } \ee
Here, $o ({\lp^2}/{a_{o}})$ denote terms which, when multiplied by
$\lp^2/a_o$ approach zero in the limit $a_o$ tends to infinity.
Thus, the leading order contribution to the entropy is indeed
proportional to the horizon area. However, even for large black
holes, one obtains agreement with the Hawking-Bekenstein formula
\emph{only} in the sector of quantum geometry in which the
Barbero-Immirzi parameter $\gamma$ takes the value $\gamma =
\gamma_o$. Thus, while all $\gamma$ sectors are equivalent
classically, the standard quantum field theory in curved
space-times is recovered in the semi-classical theory only in the
$\gamma_o$ sector of quantum geometry. It is noteworthy that
thermodynamic considerations involving \textit{large} black holes
can be used to fix the quantization ambiguity which dictates such
Planck scale properties as eigenvalues of geometric operators. As
noted in section \ref{s1}, the value of $\gamma$ can be fixed by
demanding agreement with the semi-classical result just in one
case ---e.g., a spherical horizon with zero charge, or a
cosmological horizon in the de Sitter space-time, or, \ldots .
Once the value of $\gamma$ is fixed, the theory is completely
determined and in that theory, agreement with the
Bekenstein-Hawking result holds for \textit{all} isolated horizons
with minimally coupled matter fields.

\section{Non-minimal coupling}
\label{s3}

Let us now turn to the non-minimally coupled scalar field
discussed in \cite{acs2}. Since we are considering type I
horizons, the scalar field $\phi$ is time-independent and
spherically symmetric --- hence, constant--- on $\Delta$. Here, we
wish to focus on the sector of the theory in which the inner
boundary is a type I isolated horizon with area $a_o$ and scalar
field $\phi_o$.

The detailed analysis of \cite{acs2} is based on a first order
action. One can carry out a Legendre transformation of that action
and pass to the real canonical variables analogous to those used
in section \ref{s2} by a $\gamma$-dependent canonical
transformation. However, because of the non-minimal coupling, the
symplectic structure is now modified. In place of (\ref{sym1}) we
have%
\footnote{At first it may appear that there is a discrepancy of a
factor of $-2$ in the first term of this symplectic structure and
the one in \cite{acs2}. Note, however, that in \cite{acs2} trace
is performed in the Lie algebra of the Lorentz group while here
the group is $\SU(2)$ and the $-2$ arises from the relation
between the two.}
\ba \label{sym2} {\bf \Omega}(\delta_1, \delta_2) \, =\!\!&&
\f{1}{8\pi G}  \int_M \, {\rm  Tr}\, \left[\delta_1 A \wedge
\delta_2 (f (\phi) \Sigma) - \delta_2 A \wedge \delta_1 (f(\phi)
\Sigma)\right] \nonumber\\
&+& \int_M\ K(\phi) \left[\delta_1 \phi\, \delta_2( {}^\star
 \d\phi) - \delta_2 \phi\, \delta_1({}^\star \d\phi)
 \right]\,\,
+\, \f{a_o f(\phi_o)}{\gamma\, \pi}\, \oint_S \, \left[\delta_1 W
\wedge \delta_2 W\,  \right]\, , \ea
where $f(\phi)$ is the function responsible for the non-minimal
coupling in the action (\ref{action}) and $K(\phi)$ is an
algebraic function of $\phi$, given by:
\be \label{K} K(y) = [1 + (3/16\pi G) (f'(y))^2/f(y)]\, . \ee
The classical analysis requires that $f(\phi)$ be non-zero in a
neighborhood of $S$ and of infinity and for definiteness we will
assume that it is positive there. The form of terms in
(\ref{sym2}) has two interesting implications. First, the form of
the gravitational bulk term tells us that the momentum
$\Pi^{i}_{ab}$ conjugate to the gravitational connection is given
by
\be \label{pi} \Pi^{i}_{ab} \, = \, f(\phi) \Sigma^{i}_{ab}\, ,\ee
whence the orthonormal triad $E^a_i$ of density weight $1$ is now
given by
\be \label{triad} E^{ai} = \gamma\, [f(\phi)]^{-1} \, \eta^{abc}
\Pi_{bc}^i \ee
Thus, \emph{the Riemannian geometry of $M$ is no longer dictated
just by the momentum canonically conjugate to the gravitational
connection but depends also on the scalar field}. This is a
striking, qualitative difference from the case when one has only
minimally coupled matter fields. Next, let us consider the surface
term. Since it does not contain variations of $\phi$, there is no
surface symplectic structure for the scalar field whence, as in
section \ref{s2}, the surface Hilbert space will continue to
describe quantum states only of the horizon geometry. However, a
major difference is that the coefficient of the surface term now
involves the value $\phi_o$ of the scalar field on the horizon.
Consequently, \emph{the quantum horizon geometry will now depend
on $\phi_o$}.

The total kinematic Hilbert space $\H$ again has the form $\H =
\H_V \otimes \H_S$. States in the volume Hilbert space $\H_V$ now
describe not only the polymer excitations of the geometry but also
the excitations of the scalar field (which reside at vertices of
graphs at which the geometry is excited) \cite{tt,als}. Because of
the form of the surface term in the symplectic structure, the
surface Hilbert is again the space of states of the $\U(1)$
Chern-Simons theory on the punctured $S$ (with an arbitrary number
of punctures). The level, which is dictated by the coefficient of
the surface term in (\ref{sym2}), is given by
\be \label{level2} k = \f{a_o \,f(\phi_o)}{4\pi\gamma\lp^2}\, ;
\ee
it now depends on the horizon value of the scalar field on $S$.

Next, let us consider the horizon boundary condition. In terms of
geometric fields, it is again (\ref{bc1}). However, to promote it
to the quantum theory, we need to first express it in terms of the
momentum conjugate to $A_a^i$ and therefore now depends also on
the scalar field:
\be \label{bc2}  F :\=\,\, \d W \,\,\=\,  -
\f{2\pi\g}{a_o\,f(\phi_o)}\,\, \underline\Pi^i\, r_i .\ee
To encode in the quantum theory the fact that $\IH$ is a type I
isolated horizon with area $a_o$ and scalar field $\phi_o$, we now
demand that states $\Psi$ must satisfy
\be \label{qbc2} (1\otimes \hat{F})\, \Psi \,=\,  -
\left(\frac{2\pi\gamma}{a_{o} f(\phi_o)}\, (\hat{\underline{\Pi}}
\cdot r)\otimes 1\right)\, \Psi \, . \ee
As before, a `sufficient number' of solutions exist if and only if
the bulk and the surface operators in this condition have a large
number of common eigenvalues. Since $\Pi^i_{ab}$ is the momentum
conjugate to $A_a^i$, its eigenvalues can be read off from bulk
quantum geometry and are the same as before, whence in place of
(\ref{triadev1}), the eigenvalues of the bulk operator are now
given by:
\be \label{triadevs2}
-\,\left(\f{2\pi\,\gamma}{a_{o}f(\phi_o)}\right)\, \left(8\pi
\lp^2 \, \sum_I\, m_I\, \delta^3(x, p_I)\, \eta_{ab} \right)\, ,
\ee
The eigenvalues of $\hat{F}$ are dictated by the `level' $k$ of
the Chern-Simons theory, which is now given by (\ref{level2}).
Therefore, the eigenvalues of the surface operator are given by:
\be \label{Fev2} \f{2\pi}{k}\, \sum_I\, n_I\, \delta^3(x, p_I)\,
 \equiv \,  2\pi\, \f{4\pi \gamma\lp^2}{a_of(\phi_o)}\,
\sum_I\, n_I \,\delta^3(x, p_I)\, \ee
Thus, again the eigenvalues agree when $n_I$ are integers modulo
$k$: Although the scalar field enters the quantum geometry
operators in the bulk and the level of the surface Chern Simons
theory because of non-minimal coupling, the two effects compensate
each other precisely and the delicate balance between the volume
and the surface theories required for the emergence of a coherent
description of the quantum horizon persists.

Using this description, we can calculate entropy as before. The
main differences are: i) we have to incorporate the scalar field
in the construction of the micro-canonical ensemble, and ii) now
the area operator is built from the gravitational momentum
\emph{and} the scalar field.

Starting from the expression of the area function $a_S$ on the
classical phase space and noting that $f(\phi)$ is a constant,
$f(\phi_o)$, when restricted to $S$, we can repeat the procedure
of \cite{al4} to introduce an area operator $\hat{A}_S$ on $\H_V$.
Its eigenvalues are now given by:
\be \label{ev2} \f{8\pi \gamma\lp^2}{f(\phi_o)}\,\, \sum_I
\sqrt{j_I(j_I +1)}\, . \ee
where $j_I$ are half integers and, as before, $I$ label punctures
made by the polymer excitations at the horizon. Let us  denote by
$\H_{V, a_o,\epsilon}^{\rm geometry}$ the subspace of the geometry
states in the bulk spanned by eigenvectors of the area operator
$\hat{A}_S$ with eigenvalues
\be \label{micro2} a_o - \epsilon \le \f{8\pi
\gamma\lp^2}{f(\phi_o)}\,\, \sum_I \sqrt{j_I(j_I +1)}\, \le a_o
+\epsilon\, . \ee

Next, let us consider the bulk states of the scalar field. In the
polymer framework \cite{tt,als}, the quantum scalar field resides
at vertices of graphs and can take continuous values at each
vertex. Since the scalar field takes the value $\phi_o$ on $S$
\emph{on the entire phase space}, we will restrict ourselves to
those bulk states which are eigenvectors of the scalar field
operators $\hat{\phi}(p_I)$ associated with the punctures $p_I$ on
$S$ where the eigenvalues lie in a small interval%
\footnote{$\epsilon^\prime$ is distinct from $\epsilon$ because
whereas the spectrum of the area operator is discrete, that of the
scalar field operator is continuous. Physical considerations suggest
that $\epsilon^\prime$ be constrained through: $0< \epsilon^\prime $
and $|f(\phi_o\pm \epsilon^\prime)-f(\phi_o)| < 8\sqrt{3}\pi \gamma
\lp^2/a_o$. }
around $\phi_o$:
\be \label{micro3} \phi_o -\epsilon^\prime \le\,\,\, \phi_I \equiv
\phi(p_I) \,\,\,\le \phi_o + \epsilon^\prime \, .\ee
Denote this sub-space by $\H_{V,\phi_o, \epsilon^\prime}^{\rm
scalar}$. Since there is no surface term in the symplectic
structure for the scalar field, there is no surface Hilbert space
for the scalar field. However, since the `level' of the
Chern-Simons theory depends on the value $\phi_o$ of the scalar
field at the horizon, the surface Hilbert space of geometry now
depends on $\phi_o$.

Using these notions, we can now construct the micro-canonical
ensemble. It consists of states in $\H = \H_V \otimes \H_S$ which:
i) satisfy the quantum horizon boundary conditions; ii) for which
the `volume part' of the states lies in the subspace $\H^{\rm
geometry}_{V,a_o,\epsilon} \otimes \H^{\rm
scalar}_{V,\phi_o,\epsilon'}$; and, iii) which satisfy the quantum
Einstein's equations. Thus, the overall procedure is the same as
in the minimally coupled case. For reasons explained in detail in
\cite{abk}, the entropy is given by the logarithm of the number
${\cal N}$ of \emph{surface states} in this ensemble in the sense
described below.

Quantum Einstein's equations can be imposed following the same
procedure as in the minimally coupled case. As before, these are a
set of three constraints. The implementation of the Gauss and the
diffeomorphism constraints is the same as in \cite{abk}. The first
says that the `total' state in $\H$ be invariant under the
$\SU(2)$ gauge rotations of triads and, as in \cite{abk}, this
condition is automatically met when the state satisfies the
quantum boundary condition (\ref{qbc2}). The second constraint
says that two states are physically the same if they are related
by a diffeomorphism. The detailed implementation of this condition
is rather subtle because an extra structure is needed in the
construction of the surface Hilbert space and the effect of
diffeomorphisms on this structure has to be handled carefully
\cite{abk}. However, as in the minimally coupled case, the final
result is rather simple: For surface states, what matters is only
the number of punctures; their location is irrelevant. The last
quantum constraint is the Hamiltonian one. As in the non-minimally
coupled case, in the classical theory, the constraint is
differentiable on the phase space only if the lapse goes to zero
on the boundary. Therefore, this constraint restricts only the
volume states.  However, as in the minimally coupled case, there
is an indirect restriction on surface states which arises as
follows. Consider a set $(p_I, j_I)$ with $I= 1,2,\ldots N$
consisting of $N$ punctures $p_I$ and half-integers $j_I$ real
satisfying (\ref{micro2}). We will refer to this set as `surface
data'. Suppose there exists a bulk state satisfying the
Hamiltonian constraint which is compatible with this `surface
data'  and \textit{some} choice of real numbers $\phi_I$
satisfying (\ref{micro3}). Then, we can find compatible surface
states such that the resulting states in the total Hilbert space
$\H$ lie in our ensemble. The space $S_{(p_I,j_I)}$ of these
surface states is determined entirely by the surface data. In our
state counting, we include the number ${\cal N}_{(p_I,j_I)}$
of these surface states.%
\footnote{Note that there may be a large number --possibly
infinite-- of bulk states which are compatible with a given
`surface data' in this sense. This number does not matter because
the bulk states are `traced out' in calculating the entropy of the
horizon. What matters for the entropy calculation is only the
dimensionality of $S_{(p_I, j_I)}$.}
If, on the other hand, there is no bulk state satisfying the
Hamiltonian constraint which is compatible with this `surface
data', then the surface states in $S_{(p_I,j_I)}$ will not appear
in our ensemble and will be excluded in the counting. The total
number ${\cal N}$ of states responsible for the black hole entropy
is obtained by adding up ${\cal N}_{(p_I,j_I)}$ corresponding to
each $S_{(p_I,j_I)}$ admitted in our ensemble.

As in the minimally coupled case, we now have to make a mild
assumption: we will assume that \emph{given generic surface data,
there is at least one bulk state which satisfies the Hamiltonian
constraint for some choice of $\phi_I$ satisfying (\ref{micro3}).}
Under this assumption ---on which we comment below--- the counting
can be done as in \cite{abk}, and one finds:
\be \label{S4} S_{\IH} := \ln {\cal N} = \frac{\gamma_o}{\gamma}\,
\frac{f(\phi_o)a_{o}}{4\,\lp^2} + o
\left(\frac{\lp^2}{a_{o}}\right) ,\quad {\rm where} \quad \gamma_o
= \frac{\ln 2}{\pi \sqrt{3}} \ee
where, as before, $o ({\lp^2}/{a_{o}})$ denote terms which, when
multiplied by $\lp^2/a_o$ approach zero as $a_o$ tends to
infinity. Thus, the entropy now depends on the scalar field and
for isolated horizons with large $a_o$ one recovers the
classically expected expression (\ref{S2}) using \emph{the same
value} $\gamma_o$ of the Barbero Immirzi parameter as in the
minimally coupled case.

We will conclude with a few remarks.

1. Even though there is no complete quantum gravity theory, the
calculation of black hole entropy has been possible in certain
approaches because one can encase all the difficult issues
pertaining to full quantum dynamics in a plausible assumption. In
string theory, one assumes that non-perturbative effects such as
the interactions between branes and anti-branes can be neglected;
in the symmetry based approaches a la Carlip one assumes that
certain transformations will be admissible symmetries of the full
quantum theory; and, in quantum geometry one assumes that, for
generic `surface data', the bulk Hamiltonian constraint will admit
at least one solution. However, there is a difference between the
minimally coupled fields discussed in \cite{abk} and non-minimal
couplings considered here. In the minimally coupled case the
assumption was that for each choice of the set $\{j_I\}$ such that
the \emph{total} area lies in the range $(a_o-\epsilon,
a_o+\epsilon)$, there is a bulk state satisfying the Hamiltonian
constraint. Now, we assume that a solution to the (coupled)
Hamiltonian constraint exists for some choice of the boundary
values $\phi_I$ of $\phi$, where \emph{each} $\phi_I$ is in the
range $(\phi_o -\epsilon^\prime, \phi_o+\epsilon^\prime)$. Is this
too stringent a requirement? Should one instead assume only that
an `average' of the $\phi_I$ (perhaps weighted suitably by the
$j_I$) equal $\phi_o$? It is clear that our requirement is a
better quantum representation of the fact that, classically,
$\phi\mid_S =\phi_o$ on the \emph{entire} phase space. However,
would the Hamiltonian constraint admit solutions where \emph{at
each puncture} $\phi_I$ is close to $\phi_o$? There are two
reasons suggesting that the answer is in the affirmative. First,
since we have the same number of constraints but more fields now,
it should be easier to satisfy the quantum constraints in the
bulk. This is certainly the case for classical constraints.
Second, since the spectrum of $\hat{\phi}$ is continuous, by
allowing $\phi_I$ to lie in the interval $(\phi_o-\epsilon^\prime,
\phi_o+\epsilon^\prime)$ we are letting the bulk state to lie in
an \emph{infinite} dimensional sub-space of $\H_{V,\phi}$ at each
puncture. Therefore, our assumption on the existence of solutions
to the Hamiltonian constraints seems to be rather weak. Indeed, a
priori, because of the first point, it seems weaker than that in
the minimally coupled case. To summarize, it seems plausible to
assume that isolated horizons with large areas, satisfying the
field equations and the condition $\phi\mid_S = \phi_o$ of our
classical phase space will be modelled by states in our
micro-canonical ensemble.

2. There are two `polymer representations' for the scalar field.
In the first $\hat{\phi}(x)$ has well defined action, admits any
real number as an eigenvalue and  all its eigenvectors are
normalizable. In the second representation, $\hat\pi(x)$, the
field canonically conjugate to $\hat{\phi}$, has these nice
properties but $\hat{\phi}(x)$ are not well-defined; only
$\widehat{\exp({i\mu\phi}}(x))$ are well-defined for arbitrary
real numbers $\mu$ \cite{als}. The first representation is better
suited in cases where the scalar field is non-minimally coupled
and/or has non-trivial potentials. (The situation is similar in
quantum mechanics, where the `polymer particle representation' in
which the position operator has nice properties is better suited
to deal with systems with general potentials \cite{afw}.) If one
chooses the second representation, the definition of the sub-space
$\H_{V,\phi_o,\epsilon'}^{\rm scalar}$ in the construction of the
micro-canonical ensemble has to be modified suitably.  These
modifications are technically difficult but will not affect the
final result.

3. It has been recently suggested \cite{od} that one should use
${\rm SO(3)}$ rather than $\SU(2)$ as the group of internal
rotations. In this case, to recover the Bekenstein-Hawking formula
(\ref{S1}), the Barbero Immirzi parameter $\gamma$ has to be set
equal to $\gamma_o^\prime = \ln 3/(2\pi\sqrt{2})$ in the minimally
coupled case. Our calculation would then show that, in the
non-minimally coupled case, the correctly modified expression
(\ref{S2}) results for the \emph{same} value $\gamma_o^\prime$ of
$\gamma$. However, since the motivation behind the suggestion is
somewhat ad-hoc and fails to be robust \cite{pss}, this remark is
meant only to be a mathematical observation.

4. In this paper, we restricted ourselves to type I horizons
because we wanted to focus only on the modifications introduced by
non-minimal coupling. Type II horizons can be included by
combining this analysis with that of \cite{aaentropy}. The main
modification is that for type II horizons, one has to define the
micro-canonical ensemble by specifying not just a constant
$\phi_o$ but all the (invariantly defined) multipoles of $\phi$ at
the horizon. This extension and technical issues contained in the
foregoing remarks will be discussed elsewhere.

\section{Discussion}
\label{s4}

Let us summarize. We found that in presence of non-minimal
couplings, quantum geometry in the bulk undergoes a qualitative
change because the triads, which dictate the Riemannian geometry
in the bulk, now depend not just on the gravitational momentum
$\hat{\Pi}^i$ but also on the scalar field $\hat{\phi}$.
Eigenvalues of $\hat{\Pi}^i$ are discrete, in fact the same as
those of the gravitational momentum $\hat{\Sigma}^i$ in the
minimally coupled case. But the spectrum of $\hat\phi$ is
continuous. Consequently, the kinematic arena provided by the bulk
quantum geometry is significantly different in the Einstein frame
from that in the Jordan frame. However, simplifications arise at
the horizon because, \emph{on the entire phase space}, the scalar
field takes a fixed, constant value $\phi_o$ there. In particular,
the triad operator smeared on $S$ is simply rescaled by
$f(\phi_o)$. Similarly, the only essential difference for the
surface Hilbert space is that the expression of the level of the
Chern-Simons theory is rescaled (from (\ref{level1}) to
(\ref{level2})). Consequently, the delicate matching between the
bulk and surface theories required by the quantum boundary
condition continues to be met. Using the same value (\ref{bi}) of
the Barbero Immirzi parameter as in the minimally coupled case,
for large black holes the statistical mechanical entropy is now
given by $S = (f(\phi_o)\, a_o/ 4\,\lp^2)$ in the Jordan frame
under consideration. In the case when $f(\phi)$ is everywhere
positive on $M$, we could also have worked in the Einstein frame
and the isolated horizon conditions would have been met again.
Then, the analysis of \cite{abk} would have led us to the
expression $\bar{S} = \bar{a}_o/4\,\lp^2$, where $\bar{a}_o$ is
the horizon area in the Einstein frame. However, from the relation
between the two frames it follows that $\bar{a}_o = f(\phi_o)a_o$,
whence the numerical value of entropy would be the same.

The calculations involved in our analysis are rather
straightforward. Basically, there is now a new constant
$f(\phi_o)$ and we have to keep track of how it modifies the
analysis of \cite{abk}. However, the underlying conceptual issues
are interesting. First, a priori, it was not clear that the
Chern-Simons form of the surface symplectic structure would be
preserved and the only modification would be in the expression of
the level $k$. Secondly, because the action is now significantly
different, it is far from being obvious that the geometrical,
horizon boundary condition can be expressed in terms of the
\emph{canonical variables} a controllable fashion. It is a
pleasant surprise that it \emph{can be} so expressed and,
furthermore, the modifications precisely compensate each other so
that the quantum horizon condition continues to have solutions.
Quantum geometry, developed in the mid-nineties [9-22], was based
on the Hamiltonian framework where triads are canonically
conjugate to the gravitational connection and Riemannian geometry
can be built from \textit{only} the gravitational sector of the
phase space. From this perspective, the appearance of the scalar
field in the expression of triads is a qualitative change. It
could well have happened that the rather delicate matching
required for a coherent theory of the quantum horizon falls apart
in Jordan frames. The fact that the non-minimal coupling can be
naturally accommodated, and the analysis leads to the entropy
expression suggested classically by the first law, shows that the
framework is robust.

\begin{acknowledgments}
We would like to thank Bob Wald for suggesting that non-minimal
couplings offer an interesting test for the black hole entropy
calculations in loop quantum gravity, Rob Myers, Hanno Sahlmann
and Daniel Sudarsky for discussions and the referee for comments
which significantly improved the presentation. This work was
supported in part by the NSF grants PHY-0090091, CONACyT grant
J32754-E and DGAPA-UNAM grant 112401, the Eberly research funds of
Penn State, and the Schr\"odinger Institute in Vienna.
\end{acknowledgments}

\end{document}